\documentclass[conference]{IEEEtran}
\IEEEoverridecommandlockouts
\usepackage{cite}
\usepackage{amsmath,amssymb,amsfonts}
\usepackage{algorithmic}
\usepackage{graphicx}
\usepackage{textcomp}
\usepackage{hyperref}
\ifCLASSOPTIONcompsoc
\usepackage[caption=false,font=normalsize,labelfon
t=sf,textfont=sf]{subfig}
\else
\usepackage[caption=false,font=footnotesize]{subfi
g}
\fi
\usepackage{xcolor}
\usepackage{lipsum}

\def\BibTeX{{\rm B\kern-.05em{\sc i\kern-.025em b}\kern-.08em
    T\kern-.1667em\lower.7ex\hbox{E}\kern-.125emX}}

\makeatletter % Needed to access @commands
\def\ps@IEEEtitlepagestyle{% % Redefine the page style for the title page
  \def\@oddfoot{\mycopyrightnotice}% Footer for odd pages (first page is odd)
  \def\@evenfoot{}% % Footer for even pages (empty if not needed)
}
\def\mycopyrightnotice{%
  {}
  \gdef\mycopyrightnotice{}% % This ensures it only appears on the first page
}
\makeatother
    
\begin{document}

\makeatletter
\newcommand\singlelipsum[1]{%
  \begingroup\let\lips@par\relax\csname lipsum@\@roman{#1}\endcsname
\endgroup }
\makeatother

\title{QuMeld: A Modular Framework for Benchmarking Qubit Mapping Algorithms\\ }

\author{\IEEEauthorblockN{Gabrielius Keibas}
\IEEEauthorblockA{\textit{Institute of Computer Science} \\
\textit{Vilnius University}\\
Vilnius, Lithuania \\
gabrielius.keibas@mif.stud.vu.lt}
\and
\IEEEauthorblockN{Linas Petkevičius}
\IEEEauthorblockA{\textit{Institute of Computer Science} \\
\textit{Vilnius University}\\
Vilnius, Lithuania \\
linas.petkevicius@mif.vu.lt}
}
\maketitle
\IEEEpubidadjcol

\begin{abstract}
The qubit mapping problem is a challenge in quantum computing that is related to mapping logical qubits to the physical ones on the quantum computer. Due to the diversity of quantum computer topologies and circuits, numerous approaches solving this problem exist. Finding the best solution for specific combination of topology and circuit remains difficult and no unified framework currently exists for systematically evaluating and comparing qubit mapping algorithms across different cases. We present QuMeld, an open-source framework that is designed for solving this issue. The framework currently supports six qubit mapping algorithms, sixteen quantum computer topologies and multiple evaluation metrics. The modular design of the framework allows integration of new mapping algorithms, quantum circuits, hardware topologies, and evaluation metrics, ensuring extensibility and adaptability to future developments.
\end{abstract}


\section{Introduction}

Quantum computing is rapidly advancing technology that is used in chemistry \cite{cao2019quantum}, optimisation \cite{farhi2014quantum}, machine learning \cite{cao2020quantum}. The current generation of quantum computers is called Noisy Intermediate-Scale Quantum (NISQ) era \cite{Preskill2018quantumcomputingin}. Devices have 50-300 qubits that are effected by significant noise and limited connectivity between qubits. These physical limitations create a challenge: quantum algorithms are typically designed assuming ideal, fully-connected qubit topologies. However, real quantum computers topologies enforces strict connectivity constraints allowing each qubit to interact directly only with a subset of qubits that are the neighbours.

\subsection{Qubit mapping problem}

The \textit{qubit mapping problem} arises when executing multi-qubit quantum gates on physical quantum computers. When a quantum algorithm requires a two-qubit gate (such as CNOT) between qubits that are not physically connected in the hardware topology, the circuit must be transformed by inserting additional gates to move quantum states between adjacent qubits. This transformation must preserve the algorithm's correctness while minimising overhead in terms of gate count, circuit depth, and execution time. All of these directly impact the fidelity of NISQ era quantum computations.

Formally, given a quantum circuit $C$ with $n$ logical qubits and a quantum computer with coupling graph $G = (V, E)$ where $|V| \geq n$, the qubit mapping problem seeks a mapping $\pi: \{0, \ldots, n-1\} \to V$ and a sequence of SWAP gates such that all two-qubit gates in $C$ can be executed on edges in $E$, while optimising metrics such as:
\begin{itemize}
    \item SWAP count
    \item Two-qubit gates count
    \item Circuit depth
    \item Execution time
\end{itemize}

\subsection{Proliferation of qubit mapping algorithms}

The complexity and importance of the qubit mapping problem have led to a proliferation of proposed solutions, each with different strengths and trade-offs:

\begin{itemize}
    \item heuristic search - SABRE \cite{li2019tackling} and its variants (LightSABRE \cite{zou2024lightsabrelightweightenhancedsabre}) use bidirectional heuristic search with lookahead to greedily select SWAP placements.
    \item synthesis-first - Rustiq \cite{debrugiere2024faster} employs Clifford circuit synthesis optimised for Pauli-based quantum algorithms.
    \item time-aware routing - Doustra \cite{cheng_robust_2024} considers gate timing and qubit occupancy using dual-source shortest path algorithms.
    \item Pauli strings synthesis - PauliForest \cite{y_li_pauliforest_2024} uses feature matrices to map qubits based on Pauli string interactions.
    \item machine learning - Qiskit AI \cite{dubal2025paulinetworkcircuitsynthesis, kremer2025practicalefficientquantumcircuit} applies reinforcement learning to predict optimal transpilation strategies.
\end{itemize}

Despite this diversity, selecting the appropriate algorithm for a given combination of quantum circuit, hardware topology, and optimisation objective remains challenging. Selecting the most suited algorithm requires analysis of the selected quantum computer topology and quantum circuit. 

\subsection{Motivation for QuMeld}

In typical use cases, users rely on the default transpilation method that is provided by the quantum framework. However, this can lead to the unexpected results as the transpiled circuit may contain a lot more gates compared to the original one. For this reason, the researchers are required to find the best algorithm by themselves. This can be a challenging task due to the inconsistent evaluation, difficulty of implementing new mapping algorithms and it requires deep knowledge of the quantum framework itself. To address these issues, we created the QuMeld framework. It is designed to provided systematic evaluation and comparison of the qubit mapping algorithms. Proposed framework allows practicians to investigate and select preferable topologies before trying real computations \cite{keibas2025optimising}.

QuMeld provides:

\begin{itemize}
    \item modular architecture - clean separation between algorithms, circuits, topologies, and evaluation metrics
    \item easy extensibility - well-defined interfaces for adding new components
    \item comprehensive coverage - support for 6 state-of-the-art algorithms, 16 hardware topologies, and 6 benchmark circuits
    \item automated evaluation - consistent metric collection across all experiments
    \item open source -  available at \url{https://github.com/gabkeib/qumeld}
\end{itemize}

The remainder of this paper is organised as follows: section II discusses related work; section III presents QuMeld's architecture and design; section IV describes the framework's capabilities; and section V concludes with future directions.

\section{Related work}

\subsection{Quantum compilation frameworks}

Several comprehensive quantum computing frameworks provide circuit optimisation capabilities:

Qiskit \cite{Qiskit} is IBM's open-source quantum computing framework. It includes a sophisticated transpiler with multiple optimisation levels (0-3) and supports various routing algorithms including SABRE, AI transpiler. The transpiler is tightly integrated with IBM Quantum hardware and provides excellent support for executing circuits on real devices. However, Qiskit's set of possible algorithms might not be enough and there is no automatic suggestions of the best possible algorithm for the specific setup.

Cirq and tket provides similar functionality to Qiskit. Cirq focuses on Google's quantum hardware and grid-based topologies, while tket emphasizes hardware-agnostic optimisation with support for multiple backends. Both frameworks have a proper functionality at preparing circuits for execution but lack dedicated infrastructure for systematic algorithm comparison.

MQT QMAP \cite{wille_mqt_2023} uses SAT/SMT solvers to find provably optimal mappings for small circuits ($\leq$ 8 qubits). While this provides valuable optimal solutions as benchmarks, the NP-hard complexity limits scalability to modern quantum computers with 100+ qubits.

\subsection{Specialised mapping tools}

Several tools focus specifically on the qubit mapping problem:

QMAP and related optimal mapping tools \cite{peham_optimal_2023} partition circuits into sub-architectures to find optimal solutions. These approaches are valuable for understanding theoretical bounds but face scalability challenges for large-scale quantum computers.

Algorithm-specific optimisers have emerged for particular quantum algorithms. AOQMAP \cite{ji2024improving} targets QAOA circuits using Suzuki-Trotter decomposition, achieving significant gate reductions but requiring hardware-specific templates. Similarly, various VQE-optimised mappers \cite{ji2024algorithmorientedqubitmappingvariational} exploit the structure of variational algorithms.

\subsection{Machine learning approaches}

Recent work has applied machine learning to qubit mapping:

GNAQC \cite{t_lecompte_machine-learning-based_2023} uses graph neural networks to predict high-quality initial mappings, improving fidelity by 12.7\% over random initialization. SA-DQN \cite{li_optimization_2024} combines deep Q-networks with simulated annealing. While promising, these methods take a longer time, requires training and data to train on, proper selection of the hyper parameters.

\subsection{Positioning of QuMeld}

QuMeld is designed as a benchmarking framework for qubit mapping algorithms. While platforms like Qiskit, Cirq primarily target circuit compilation and hardware deployment, QuMeld focuses on systematic evaluation and comparative analysis of qubit mapping algorithms. Various algorithms can be accessed under a common interface and standardised benchmarks are provided. Not only benchmarking but the selection of the best algorithm can be found using this tool. Through its modular architecture, the framework allows integration of new algorithms, hardware topologies.

\section{QuMeld architecture and design}

\subsection{High-level architecture}

\begin{figure}[!t]
\centering
\includegraphics[width=\columnwidth]{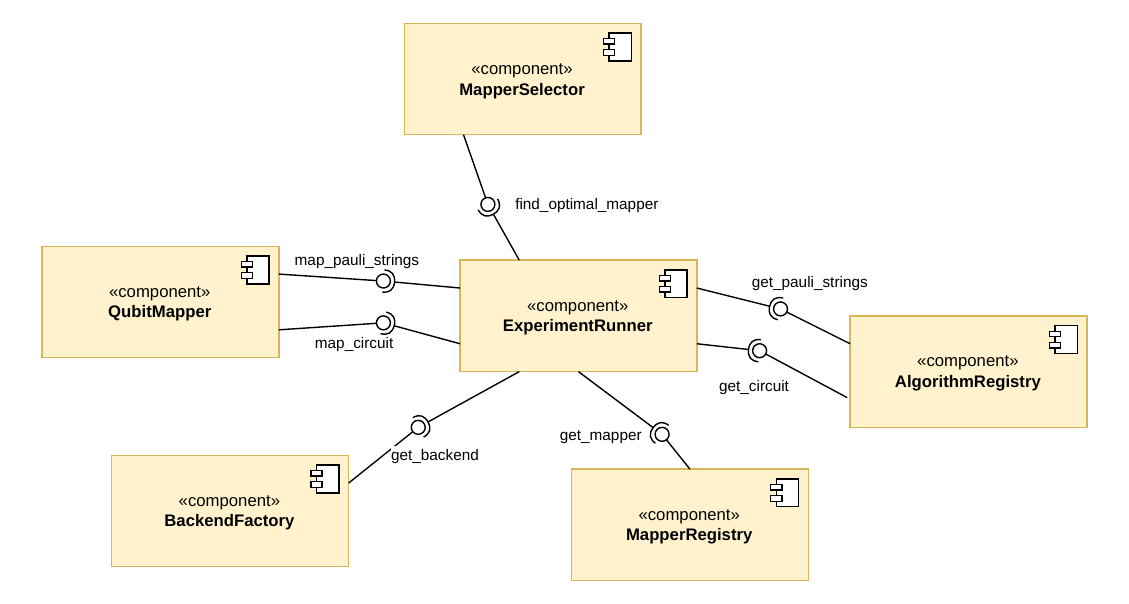}
\caption{QuMeld high level component diagram depicting interaction between mappers, circuits, topologies and results}
\label{fig:component_diagram}
\end{figure}

QuMeld high-level component diagram is presented in Figure \ref{fig:component_diagram}. \texttt{ExperimentRunner} is the central orchestrator. This design minimises coupling between components and provides a single point of coordination for experiments.

\subsection{Core components}

\subsubsection{QubitMapper interface}

The \texttt{QubitMapper} abstract base class defines two key methods:

\begin{itemize}
    \item \texttt{run\_circuit(circuit, backend)}: maps a generic \texttt{QuantumCircuit} to the topology defined by \texttt{backend}.
    \item \texttt{run\_hamiltonian(pauli\_strings, backend)}: maps a list of Pauli strings (can be used for VQE or QAOA) directly.
\end{itemize}

This dual interface accommodates both circuit-based algorithms and ones that support Pauli strings synthesis. Algorithms implement one or both methods depending on their capabilities.

Each mapper returns a \texttt{CircuitOptimisationResult} containing:
\begin{itemize}
    \item Mapped circuit
    \item SWAP count
    \item CNOT count
    \item Circuit depth
    \item optimisation time
    \item Failure reasons (if applicable)
\end{itemize}

\subsubsection{MapperRegistry}

\texttt{MapperRegistry} automatically discovers and loads available mapping algorithms from the \texttt{mappers/} package. This registry pattern enables dynamic algorithm loading without hardcoding algorithm names in the core framework.

\subsubsection{MapperSelector}

\texttt{MapperSelector} identifies the optimal algorithm for a given circuit and topology by:
\begin{enumerate}
    \item retrieving all available mappers from \texttt{MapperRegistry}
    \item running each mapper on the input circuit/topology
    \item comparing results using user-specified criteria (SWAP count, depth, etc.)
    \item returning the best-performing mapper
\end{enumerate}

\subsubsection{AlgorithmProvider and HamiltonianProvider}

These provider components supply quantum circuits and Pauli strings for testing:

\begin{itemize}
    \item AlgorithmProvider: loads predefined circuits (VQE, QAOA variants) from the framework's circuit library
    \item HamiltonianProvider: provides Pauli string representations loaded from JSON files or generated programmatically (for example, QAOA Max-Cut hamiltonians)
\end{itemize}

The provider pattern decouples data sources from algorithm implementations, making it easy to add new benchmark circuits.

\subsubsection{BackendFactory}

\texttt{BackendFactory} creates Qiskit's \texttt{BackendV2} instances representing quantum computer topologies. Each backend specifies:

\begin{itemize}
    \item coupling graph (qubit connectivity)
    \item supported gate set (currently the same one is set for all)
    \item simulated noise characteristics (T1/T2 times, gate errors)
\end{itemize}

The factory pattern abstracts topology creation, allowing users to specify topologies by name (e.g., "google\_willow", "ibm\_eagle") without worrying about implementation details.

\subsubsection{ExperimentRunner}

\texttt{ExperimentRunner} orchestrates end-to-end experiments:

\begin{enumerate}
    \item load experiment configuration (algorithms, topologies, circuits)
    \item for each (algorithm, topology, circuit) combination:
    \begin{enumerate}
        \item retrieve circuit from \texttt{AlgorithmProvider}
        \item create backend from \texttt{BackendFactory}
        \item load mapper from \texttt{MapperRegistry}
        \item execute mapping
        \item collect metrics
    \end{enumerate}
    \item store results in timestamped directories
\end{enumerate}

\subsection{Implementation details}

Language and dependencies: QuMeld is implemented in Python 3.13+ with Qiskit 2.1.1 as the primary quantum computing library. The \texttt{uv} package manager handles dependencies, and the \texttt{ruff} linter enforces type safety.

Algorithm integration strategies: different algorithms require different integration approaches:

\begin{itemize}
    \item native Python - LightSABRE (decay and lookahead heuristics), Rustiq, and Qiskit AI are integrated directly via Qiskit's pass manager system.
    \item subprocess (C++) - Doustra is implemented in C++ as part of the Qsyn framework. QuMeld invokes Doustra via Python's \texttt{subprocess} module, converting circuits to OpenQASM format for communication.
    \item isolated environment - Qiskit AI requires \texttt{networkx==2.8.5}, conflicting with QuMeld's dependencies. It runs in a separate virtual environment, communicating via serialized circuit files.
\end{itemize}

\section{Framework capabilities}

\subsection{Supported qubit mapping algorithms}

QuMeld currently integrates six state-of-the-art qubit mapping algorithms:

\subsubsection{LightSABRE (decay and lookahead heuristics)}

SABRE \cite{li2019tackling} is a bidirectional heuristic search algorithm widely considered state-of-the-art. LightSABRE \cite{zou2024lightsabrelightweightenhancedsabre} improves upon SABRE with optimised heuristics and a Rust implementation. QuMeld includes two LightSABRE configurations:

\begin{enumerate}
    \item lookahead heuristic: evaluates future gate dependencies when selecting SWAPs
    \item decay heuristic: dynamically weights qubits based on swap probability
\end{enumerate}

Both use custom Qiskit pass managers with \texttt{SabreLayout} and \texttt{SabreSwap} passes configured for high-quality results (200 layout trials, 200 swap trials, 4 forward-backward iterations).

\subsubsection{Rustiq}

Rustiq \cite{debrugiere2024faster} specialises in synthesis-first compilation for Pauli-based algorithms. It constructs Clifford circuits optimised for either gate count or depth using greedy heuristics and maximum-weight matching. QuMeld integrates Rustiq via Qiskit's high-level synthesis framework, configured for depth optimisation with 400 shuffles.

\subsubsection{Doustra}

Doustra \cite{cheng_robust_2024} combines depth-first search for initial mapping with dual-source Dijkstra routing that considers both distance and qubit occupancy time. This time-aware approach reduces gate collisions and improves parallelism. QuMeld invokes Doustra's C++ implementation via subprocess, handling circuit conversion and result parsing automatically.

\subsubsection{PauliForest}

PauliForest \cite{y_li_pauliforest_2024} is a static mapping algorithm designed specifically for Pauli string operators. It builds a feature matrix from Pauli strings, identifies high-interaction qubits, and maps them to central topology locations to minimize average distance. QuMeld integrates the PauliGo implementation, converting Pauli strings to the required block format and applying post-synthesis Qiskit optimisation.

\subsubsection{Qiskit AI}

The Qiskit AI-powered transpiler \cite{dubal2025paulinetworkcircuitsynthesis, kremer2025practicalefficientquantumcircuit} uses reinforcement learning to predict optimal transpilation strategies. It partitions circuits into blocks, optimizes each block with an AI model trained on large datasets, and reassembles the result. QuMeld integrates the \texttt{AIRouting} pass with optimisation level 3 and optimised layout method.

\subsection{Supported quantum computer topologies}

QuMeld supports 16 diverse quantum computer topologies spanning different connectivity patterns and qubit counts, including various IBM topologies, Google Willow, IonQ, Rigetti Novera, Riken Fujitsu.

\begin{itemize}
    \item IBM Heavy hex (IBM's standard): Almaden (20q), Cambridge (28q), Falcon (27q), Montreal (27q), Paughkeepsie (20q), Reuschlikon (16q), Tokyo (20q), Manhattan (65q), Rochester (53q), Eagle (127q), Heron (133q)
    \item Google Willow (105q): Google's recent quantum processor with optimised connectivity and error correction
    \item IonQ Harmony (9q): Fully connected trapped-ion topology (edge density = 1.0)
    \item Rigetti Novera (9q): Square lattice topology (avg degree $\approx 3.6$)
    \item Riken Fujitsu (256q): Large-scale superconducting quantum computer
    \item Hexagonal lattice (54q): Synthetically generated hexagonal grid
\end{itemize}

All topologies are implemented as Qiskit \texttt{BackendV2} instances with randomised noise characteristics: T1 relaxation times $\sim \mathcal{U}(50, 100)$ $\mu$s, T2 dephasing times $\sim \mathcal{U}(20, 80)$ $\mu$s, qubit frequencies $\sim \mathcal{U}(4.5, 5.5)$ GHz, and realistic gate error rates.

\subsection{Benchmark quantum circuits}

QuMeld includes six benchmark circuits representing common quantum computing workloads:

\subsubsection{VQE circuits}

\begin{itemize}
    \item H$_2$ molecule (4 qubits, 30 two-qubit gates): Small molecular hamiltonian
    \item LiH molecule (12 qubits, 6023 two-qubit gates): Large-scale molecular hamiltonian from PauliForest benchmarks
    \item Random demo (9 qubits, 60 two-qubit gates): Randomly generated Pauli strings
    \item EfficientSU2 ansatz (5 qubits, 4 two-qubit gates): Parameterised variational ansatz with linear entanglement
\end{itemize}

\subsubsection{QAOA circuits}

\begin{itemize}
    \item Max-Cut hamiltonian(5 qubits, 6 two-qubit gates): Problem hamiltonian for graph partitioning
    \item QAOA ansatz (5 qubits, 6 two-qubit gates): Mixer + problem hamiltonian structure
\end{itemize}

All Pauli-based circuits are stored as JSON files with coefficients, enabling consistent evaluation across algorithms.

\subsection{Result storage and analysis}

Results are stored in timestamped directories with hierarchical structure:
\begin{verbatim}
results/
   <timestamp>/
       <topology_name>/
           <circuit_name>/
               <mapper_name>.json
\end{verbatim}

Each JSON file contains all metrics, enabling systematic post-processing. QuMeld includes Jupyter notebooks for result loading, table generation, and CSV export.

\subsection{Extensibility}

Adding new components to QuMeld is straightforward:

\begin{itemize}

    \item new mapper - implement \texttt{QubitMapper} interface, place in \texttt{mappers/} directory. The \texttt{MapperRegistry} will auto-discover it.

    \item new topology - define a coupling graph function in \texttt{BackendFactory}, register with the factory.

    \item new circuit - place a circuit generation function or Pauli string JSON file in the appropriate provider directory.

    \item new metric - extend \texttt{CircuitOptimisationResult} and add computation logic in the mapper base class.
\end{itemize}

\section{Conclusion and future work}

We presented QuMeld, an open-source modular framework for benchmarking qubit mapping algorithms. QuMeld addresses a critical gap in quantum computing research by providing a unified platform for systematically evaluating diverse mapping algorithms across varied topologies and circuits. The framework's star architecture, well-defined interfaces, and design patterns enable easy integration of new algorithms, topologies, and evaluation metrics.

QuMeld currently supports six state-of-the-art algorithms (LightSABRE variants, Rustiq, Doustra, PauliForest, Qiskit AI), sixteen quantum computer topologies (9-256 qubits), and six benchmark circuits representing VQE and QAOA workloads. The framework's modular design accommodates diverse implementation languages (Python, C++, isolated environments) and evaluation approaches.

\subsection{Availability and use cases}

QuMeld is freely available at \url{https://github.com/gabkeib/qumeld} under an open-source license. Researchers can use QuMeld to:

\begin{itemize}
    \item algorithm selection - identify the best mapper for specific circuit/topology combinations
    \item algorithm development - benchmark new mapping algorithms against state-of-the-art baselines
    \item topology analysis - understand how hardware connectivity patterns affect mapping quality
    \item reproducible research - generate consistent, comparable results across studies
\end{itemize}

\subsection{Future works}

Several directions for future development include:

\begin{enumerate}
    \item Integration of new optimisation algorithms including EffectiveQM \cite{c_luo_towards_2025}, SWin \cite{fu_effective_2023}, and tabu search \cite{jiang_qubit_2024}.
    \item Error correction support and inclusion into comparison metrics.
    \item Improve automatic algorithm selection. by using advanced decision trees or machine learning approaches to predict optimal algorithms.
    \item Adding larger circuits for the possible testing algorithms to have a more diverse results.
    \item Performance optimisation. Parallelising experiment execution and optimizing framework overhead
\end{enumerate}

\bibliography{bibliography}
\bibliographystyle{plain}

\end{document}